# Optimization of interactive evaluation of operation efficiency of the city's transport system by methods of *U*-statistics

O. D. Polishchuk[1], M. S. Yadzhak[1,2]

Laboratory of Modeling and Optimization of Complex Systems
Pidstryhach Institute for Applied Problems of Mechanics and Mathematics, National Academy of Sciences of Ukraine,
Ivan Franko Lviv National University, Lviv, Ukraine
od_polishchuk@ukr.net

**Abstract** – Methods for interactive evaluation of the functioning efficiency of the motor transport system for a large city based on the use of U-statistics methods has been formalized. To optimize this technique, effective algorithmic constructions for parallel execution of local, aggregated and forecasting evaluation of system components on modern computing means – multi-core computers, clusters, hybrid architectures and high-performance computing environments are proposed. The obtained scientific results allow for real-time evaluation of the efficiency for the city's motor transport system.

**Вступ.** Автотранспортна система (АТС) великого міста розглядається нами як мережева система із частково впорядкованим рухом потоків, а саме засобів громадського транспорту. Серед основних причин неефективного функціонування АТС можна назвати недоліки дорожньої інфраструктури (відсутність зручних розв'язок, незадовільний стан шляхів або транспортних засобів, низька пропускна здатність доріг, надмірна щільність транспортних потоків тощо) [1], а також неналежну організацію руху за допомогою світлофорів. Необхідно зазначити, що на якості роботи цієї системи та її стані позначається велика кількість негативних випадкових чинників, серед яких ДТП, катаклізми, кліматичні умови, загроза терористичних актів або військових дій [2] тощо.

Сучасні засоби GPS-моніторингу дозволяють в режимі реального часу відстежувати місцезнаходження транспортного засобу, визначати його швидкість, місце та тривалість зупинок, фіксувати перетин контрольних зон та найменші відхилення від заданого маршруту [3] тощо. На підставі цих даних можна формувати детальні звіти про пересування конкретного транспортного засобу з подальшим їх використанням як для оцінювання його стану, так і для аналізу стану і якості функціонування АТС. Зазначимо, що згадані технології дозволяють визначати щільність, інтенсивність та обсяги транспортних потоків, які пересуваються ребрами автотранспортної мережі (АТМ) [4, 5]. Аналіз таких даних, одержаних для системи громадського транспорту, надає можливість робити в режимі реального часу опосередковані, але достатньо обґрунтовані висновки про її стан і якість функціонування. Унаслідок цього можна здійснювати адекватне та своєчасне реагування на транспортні ситуації, що виникають в місті [6]. У праці [7] запропоновано ефективну методику інтерактивного оцінювання АТС великого міста. Ця методика ґрунтується на використанні методу $U$-статистик [8, 9] та поєднує процедури локального, прогностичного та агрегованого оцінювання стану і процесу функціонування складових системи різних рівнів ієрархії. Згадана методика використовує значні обсяги вхідних даних, які надходять у режимі реального часу і потребує оптимізації за часом реалізації для своєчасного прийняття необхідних рішень. Оптимізацію будемо здійснювати на підставі використання паралельних методів організації обчислень [10–14].

Мета статті – формалізація інтерактивної методики оцінювання і прогнозування стану та ефективності функціонування АТС великого міста на основі методу $U$-статистик і розроблення паралельних алгоритмів реалізації цієї методики.

**Формалізація процедури локального оцінювання.** Для уніфікації способів оцінювання руху засобів громадського транспорту (ЗГТ) на елементарних ділянках шляху та перехрестях АТМ згідно з [7] визначаємо такі сукупності:

$$\mathbf{t}_i^{a,j} = \{t_{ik}^a\}_{k=1}^j;\ \mathbf{t}_i^{h,j} = \{t_{ik}^h\}_{k=1}^j, \qquad (1)$$

у яких $t_{ik}^a = s_i / v_{ik}^a$, $t_{ik}^h = s_i / v_{ik}^h$, де $s_i$ – довжина $i$-ї елементарної ділянки ($i = \overline{1, L}$); $L$ – кількість елементарних ділянок, з яких складається АТМ міста та якими рухається громадський транспорт; $v_{ik}^a$ – середня швидкість руху на $i$-й елементарній ділянці $k$-го з початку доби ЗГТ; $v_{ik}^h$ – середня очікувана або обчислена на підставі попередніх статистичних досліджень швидкість руху на $i$-й ділянці автотранспортних засобів у момент $t_j \in [00:00, 24:00]$.

Далі обчислюються головна $U_{ij}$- та $U_{ij}^m$-, $U_{ij}^s$-, $U_{ij}^c$-статистики руху на $i$-й ділянці АТМ протягом періоду $[0, t_j]$. При цьому $U_{ij}$-статистика визначає відносне відхилення реального часу руху ЗГТ від очікуваного. За статистикою $U_{ij}^m$ оцінюють систематичну похибку або наскільки середнє арифметичне значення реальних даних відрізняється від відповідного значення очікуваних показників. $U_{ij}^s$-статистика дозволяє оцінюва-ти міру збігу очікуваних і реальних швидкостей ЗГТ. За статистикою $U_{ij}^c$ оцінюють залишкову похибку і вона дозволяє виокремити випадки, коли задовільні за першими двома $U_{ij}^m$-, $U_{ij}^s$-статистиками реальні дані взаємно компенсують похибки спостережень. Зокрема, $U_{ij}$-статистика обчислюється за формулою [7]:

$$U_{ij} = \left\| \mathbf{t}_i^{a,j} - \mathbf{t}_i^{h,j} \right\|_{R^j} / \left( \left\| \mathbf{t}_i^{a,j} \right\|_{R^j} + \left\| \mathbf{t}_i^{h,j} \right\|_{R^j} \right), \qquad (2)$$

де $\|\mathbf{t}\|_{R^j} = \sqrt{\sum_{k=1}^{j} t_k^2}$. Тут $j = \overline{1, N_i}$; $i = \overline{1, L}$, де $N_i$ – максимальна кількість ЗГТ, які рухаються $i$-ю ділянкою протягом доби.

На підставі обчислених значень згаданих вище $U$-статистик формуємо уточнені бальні оцінки їхньої поведінки протягом періоду $[0, t_j]$. При цьому зважаємо на те, що ці статистики набувають значень із проміжку $[0, 1]$ та поведінка оцінюваних характеристик руху ЗГТ є тим кращою, чим ближчими є значення $U_{ij}$-, $U_{ij}^m$- та $U_{ij}^s$-статистик до 0, а значення $U_{ij}^c$-статистики – до 1. Тому інтерактивну статистичну уточнену бальну оцінку, наприклад, $e_{U_{ij}}(\mathbf{t}_i^{a,j})$ поведінки сукупності $\mathbf{t}_i^{a,j}$ за головною статистикою $U_{ij}$ визначаємо так:

$$e_{U_{ij}}(\mathbf{t}_i^{a,j}) = \begin{cases} 5, & U_{ij} \in [0{,}00; \gamma_1] \\ 4 + 4(\gamma_2 - U_{ij}), & U_{ij} \in ]\gamma_1, \gamma_2] \\ 3 + 4(\gamma_3 - U_{ij}), & U_{ij} \in ]\gamma_2, \gamma_3] \\ 2, & U_{ij} \in ]\gamma_3; 1{,}00]. \end{cases} \qquad (3)$$

Аналогічно визначаються інтерактивні статистичні оцінки $e_{U_{ij}^m}(\mathbf{t}_i^{a,j})$ та $e_{U_{ij}^s}(\mathbf{t}_i^{a,j})$ поведінки сукупності $\mathbf{t}_i^{a,j}$ за $U_{ij}^m$- та $U_{ij}^s$-статистиками відповідно ($j = \overline{1, N_i}$; $i = \overline{1, L}$). Зауважимо, що формула для визначення інтерактивної статистичної оцінки $e_{U_{ij}^c}(\mathbf{t}_i^{a,j})$ поведінки сукупності $\mathbf{t}_i^{a,j}$ за $U_{ij}^c$-статистикою має схожу структуру [7].

Далі отримуємо узагальнений висновок $E_{ij}(\mathbf{t}_i^{a,j})$ про поведінку сукупності $\mathbf{t}_i^{a,j}$ протягом періоду $[0, t_j]$ та відповідний фінальний узагальнений висновок $E_{i,N_i}(\mathbf{t}_i^{a,N_i})$ протягом доби за усіма $U$-статистиками, використовуючи метод лінійної агрегації, тобто

$$E_{i,N_i}(\mathbf{t}_i^{a,N_i}) = (e_{U_{ij}}(\mathbf{t}_i^{a,N_i}) + e_{U_{ij}^m}(\mathbf{t}_i^{a,N_i}) + e_{U_{ij}^s}(\mathbf{t}_i^{a,N_i}) + e_{U_{ij}^c}(\mathbf{t}_i^{a,N_i}))/4, \; i = \overline{1, L}. \qquad (4)$$

Для аналізу результатів неперервного моніторингу часу перетину ЗГТ регульованих перехресть, розташованих на шляху їхнього руху, на підставі методів $U$-статистик та обчислення відповідних уточнених бальних оцінок ефективності роботи світлофорів також використовуються співвідношення (2)–(4), при цьому сукупності $\mathbf{t}_i^{a,j}$ та $\mathbf{t}_i^{h,j}$ інтерпретуються відповідно, як реальні та очікувані значення часу затримки ЗГТ перед світлофором.

**Формалізація процедури агрегованого оцінювання.** Загалом вважаємо, що АТС міста поділена на $R$ районів, які відзначаються особливостями структури та процесу функціонування системи. Структуру $r$-го району утворюють $K_r$ вузлів та $N_r^0$ ребер. Ребра утворюють $N_r^g$ груп заданої пріоритетності. Зауважимо, що структуру кожного ребра будемо визначати як послідовність елементарних ділянок автошляху, розділених світлофорами та зупинками ЗГТ. Кількість таких ділянок у $k$-му ребрі $r$-го району дорівнює $L_{r,k}$.

Розглянемо побудову агрегованих висновків про стан ребер, підмереж району та інфраструктури міста загалом. Нехай $E_{l,m}^{r,k}$ – обчислений за формулою (4) узагальнений висновок про стан $l$-ї ділянки $k$-го ребра $r$-го району АТС, отриманий унаслідок неперервного моніторингу руху ЗГТ цією ділянкою протягом $m$-ї доби ($m = \overline{1,M}$).

Узагальнений висновок $E_{l,M}^{r,k}$ про стан досліджуваної ділянки протягом $M$ діб згідно з [7] отримуємо методом нелінійної агрегації (МНА) за співвідношенням

$$E_{l,M}^{r,k} = \prod_{m=1}^{M} E_{l,m}^{r,k} / (\varepsilon_{l,M}^{r,k})^{M-1}, \tag{5}$$

де $\varepsilon_{l,M}^{r,k} = \sum_{m=1}^{M} E_{l,m}^{r,k} / M$.

Далі, аналогічно до (5), використовуючи МНА, будуємо узагальнений висновок $E_M^{r,k}$ про стан ребра АТМ, яке складається з уже оцінених $L_{r,k}$ елементарних ділянок.

Висновок $E_M^r$ про стан автошляхів району одержуємо методом гібридної агрегації (МГА):

$$E_M^r = \sum_{n=1}^{N_r^g} (\rho_n^r \widetilde{E}_M^{r,n}) / \sum_{n=1}^{N_r^g} \rho_n^r, \tag{6}$$

де $\rho_n^r$ ($n = \overline{1, N_r^g}$) – пріоритетність груп ребер; $\tilde{E}_M^{r;n}$ – узагальнений висновок для сукупності ребер *n*-ї групи, отриманий МНА.

Далі, аналогічно до (6), використовуючи МГА, будуємо узагальнений висновок $E_M$ про стан транспортної інфраструктури АТС міста з урахуванням пріоритетності *N* груп районів та узагальненого висновку для сукупності районів кожної групи, отриманого МНА.

Доцільно використовувати також прогнозування поведінки оцінок типу $E_M^{r;k}$, $E_M^r$ та $E_M$ під час планування термінів та обсягів необхідних ремонтних робіт автошляхів міста і пов'язаних з ними витрат [15].

Розглянемо побудову узагальнених висновків для сукупності локальних оцінок, які характеризують ефективність режимів роботи світлофорів на автошляхах міста. Ці висновки формуються аналогічно до розглянутих вище і передбачають виконання наступних кроків [7]:
1) нелінійна агрегація за часом (подобово) оцінок режиму роботи окремого світлофора протягом певного періоду часу;
2) побудова МНА та прогноз поведінки узагальненого висновку про ефективність роботи усіх оцінюваних на попередньому кроці світлофорів окремого регульованого перехрестя АТМ;
3) побудова на основі МГА та прогнозування поведінки узагальненого висновку про ефективність роботи усіх світлофорів району міста з урахуванням пріоритетності окремих його регульованих перехресть;
4) побудова на підставі МГА та прогнозування поведінки висновку про ефективність роботи усіх світлофорів міста з урахуванням пріоритетності окремих його районів.

Доцільним також є агреговане оцінювання для аналізу та прогнозування розвитку поточної автотранспортної ситуації, що склалася на окремому маршруті руху ЗГТ, у районі міста або його АТС загалом. При цьому враховується пріоритетність окремих ребер та світлофорів (розташованих, наприклад, на шляху заданого маршруту).

Зазначимо також, що агреговані оцінки можна будувати і для сукупності елементарних ділянок шляху та світлофорів (взятих разом), які складають АТМ міста.

**Розпаралелювання опрацювання даних під час локального оцінювання.** Для паралельного опрацювання даних під час локального оцінювання нами пропонується паралельно-послідовний підхід, який описується конструкцією (7), з використанням примітивів галуження, злиття *fork*, *join* [16].

$$fork\,(h_1^a, h_2^a, \ldots, h_{2N_\Sigma}^a)\,join$$

$$fork\,(h_1^{nor}, h_2^{nor}, \ldots, h_{3N_\Sigma}^{nor})\,join$$

$$fork\,(h_1^U, h_2^U, \ldots, h_{4N_\Sigma}^U)\,join \qquad (7)$$

$$fork\,(h_1^e, h_2^e, \ldots, h_{4N_\Sigma}^e)\,join$$

$$fork\,(h_1^E, h_2^E, \ldots, h_{N_\Sigma}^E)\,join.$$

У наведеній алгоритмічній конструкції (7) $N_\Sigma = N_1 + N_2 + \ldots + N_L$; $h_l^a$ ($l = \overline{1, 2N_\Sigma}$), $h_{l_1}^{nor}$ ($l_1 = \overline{1, 3N_\Sigma}$), $h_n^U$, $h_n^e$ ($n = \overline{1, 4N_\Sigma}$) та $h_{n_1}^E$ ($n_1 = \overline{1, N_\Sigma}$) – набори автономних паралельних гілок, в кожному з яких обчислюються відповідно сукупності $\mathbf{t}_i^{a,j}$ та $\mathbf{t}_i^{h,j}$; норми $\left\|\mathbf{t}_i^{a,j} - \mathbf{t}_i^{h,j}\right\|_{R^j}$, $\left\|\mathbf{t}_i^{a,j}\right\|_{R^j}$ та $\left\|\mathbf{t}_i^{h,j}\right\|_{R^j}$; значення статистик $U_{ij}$, $U_{ij}^m$, $U_{ij}^s$ та $U_{ij}^c$; оцінки $e_{U_{ij}}(\mathbf{t}_i^{a,j})$, $e_{U_{ij}^m}(\mathbf{t}_i^{a,j})$, $e_{U_{ij}^s}(\mathbf{t}_i^{a,j})$ та $e_{U_{ij}^c}(\mathbf{t}_i^{a,j})$ і узагальнені висновки $E_{ij}(\mathbf{t}_i^{a,j})$. При цьому слід мати на увазі, що $j = \overline{1, N_i}$; $i = \overline{1, L}$.

Зазначимо, що запропонована алгоритмічна конструкція (7) дозволяє під час локального оцінювання залучати значний обсяг паралелізму. Однак, необхідно враховувати, що існують додаткові резерви паралелізму, зокрема, під час обчислення власне норми, конкретного значення тієї чи іншої статистики, уточненої бальної оцінки (наприклад, перевірку умов можна здійснювати одночасно) та узагальненого висновку (додавання оцінок можна здійснювати в режимі повного двійкового дерева). Згадані резерви є несуттєвими, тому ми їх не розглядатимемо детальніше.

Далі оцінимо прискорення для паралельних обчислень згідно з (7). Зазначимо, що паралельні гілки в межах кожного із поданих вище наборів мають приблизно однакову складність. Припустимо, що час обчислень для кожного набору гілок дорівнює відповідно $t_1^a, t_2^{nor}, t_3^U, t_4^e, t_5^E$. Отже, час обчислень, виконуваних під час локального оцінювання в послідовному режимі, дорівнюватиме

$$(2t_1^a + 3t_2^{nor} + 4(t_3^U + t_4^e) + t_5^E)N_\Sigma, \qquad (8)$$

а в паралельному, згідно з (7), обчислюватиметься за формулою

$$t_1^a + t_2^{nor} + t_3^U + t_4^e + t_5^E. \qquad (9)$$

На підставі (8) і (9) отримуємо формулу для прискорення $S_1$ паралельних обчислень [17], виконуваних під час локального оцінювання згідно з (7):

$$S_1 = (2t_1^a + 3t_2^{nor} + 4(t_3^U + t_4^e) + t_5^E)N_\Sigma /(t_1^a + t_2^{nor} + t_3^U + t_4^e + t_5^E). \quad (10)$$

Після деяких еквівалентних перетворень із формули (10) отримуємо:

$$S_1 = (2 + (t_2^{nor} + 2t_3^U + 2t_4^e - t_5^E)/(t_1^a + t_2^{nor} + t_3^U + t_4^e + t_5^E))N_\Sigma.$$

Оцінюючи складність відповідних обчислень, легко отримати, що

$$t_2^{nor} + 2t_3^U + 2t_4^e - t_5^E > 0,$$

тобто на основі цієї нерівності одержуємо, що

$$S_1 > 2N_\Sigma.$$

Оскільки передбачається, що під час досліджень розглядається значна кількість елементарних ділянок, по кожній з яких протягом доби рухається багато ЗГТ, тобто ділянки є достатньо завантаженими, то можна очікувати, що прискорення $S_1$ паралельних обчислень буде суттєвим.

Конструкція (7) складається з автономних паралельних гілок, тому не буде виникати жодних неочікуваних проблем під час її програмної реалізації на сучасних паралельних обчислювальних системах [18], таких як багатоядерні комп'ютери та кластери, які є зараз широкодоступними засобами.

Локальне оцінювання проходження ЗГТ ділянок АТС та ефективності роботи світлофорів можна здійснювати одночасно, тобто паралельно.

**Організація паралельних обчислень під час агрегованого оцінювання.** Для ефективної реалізації процедури агрегованого оцінювання нами запропоновано алгоритмічну конструкцію (11), яка також використовує паралельно-послідовний підхід до організації обчислень.

$$\begin{aligned} &fork\,(h_1^M, h_2^M, \ldots, h_{M_1}^M)\,join \\ &fork\,(h_1^L, h_2^L, \ldots, h_{M_2}^L)\,join \quad (11) \\ &fork\,(h_1^N, h_2^N, \ldots, h_R^N)\,join \\ &h^E. \end{aligned}$$

У наведеній конструкції $h_{i_1}^M$ ($i_1 = \overline{1, M_1}$), $h_{i_2}^L$ ($i_2 = \overline{1, M_2}$), $h_{i_3}^N$ ($i_3 = \overline{1, R}$) – набори повністю автономних паралельних гілок для обчислення відповідно агрегованих оцінок

$E_{l,M}^{r,k}$ ($l = \overline{1, L_{r,k}}$; $k = \overline{1, N_r^0}$; $r = \overline{1, R}$), $E_M^{r,k}$ ($k = \overline{1, N_r^0}$; $r = \overline{1, R}$), $E_M^r$ ($r = \overline{1, R}$), а $h^E$ – фрагмент, у якому обчислюється оцінка $E_M$. При цьому $M_1 = \sum_{r=1}^{R} \sum_{k=1}^{N_r^0} L_{r,k}$; $M_2 = \sum_{r=1}^{R} N_r^0$.

Припустимо, що часи обчислень, виконуваних в гілках $h_{i_1}^M$, $h_{i_2}^L$, $h_{i_3}^N$, дорівнюють відповідно $t_{i_1}^M$, $t_{i_2}^L$, $t_{i_3}^N$ ($i_1 = \overline{1, M_1}$; $i_2 = \overline{1, M_2}$; $i_3 = \overline{1, R}$), а час обчислення у фрагменті $h^E$ дорівнює $t^0$. Унаслідок цього одержимо, що час обчислень, виконуваних під час агрегованого оцінювання в послідовному режимі буде дорівнювати

$$\sum_{i_1=1}^{M_1} t_{i_1}^M + \sum_{i_2=1}^{M_2} t_{i_2}^L + \sum_{i_3=1}^{R} t_{i_3}^N + t^0, \quad (12)$$

а у паралельному, використовуючи (11), обчислюватиметься за формулою

$$t_{\max}^M + t_{\max}^L + t_{\max}^N + t^0, \quad (13)$$

де $t_{\max}^M = \max\{t_1^M, t_2^M, ..., t_{M_1}^M\}$; $t_{\max}^L = \max\{t_1^L, t_2^L, ..., t_{M_2}^L\}$; $t_{\max}^N = \max\{t_1^N, t_2^N, ..., t_R^N\}$.

На основі оцінок часу (12) та (13) отримуємо формулу для обчислення прискорення паралельних обчислень $S_2$:

$$S_2 = \left( \sum_{i_1=1}^{M_1} t_{i_1}^M + \sum_{i_2=1}^{M_2} t_{i_2}^L + \sum_{i_3=1}^{R} t_{i_3}^N + t^0 \right) \Big/ \left( t_{\max}^M + t_{\max}^L + t_{\max}^N + t^0 \right). \quad (14)$$

Найбільш ефективним є розпаралелювання згідно (11) у випадку, коли складності паралельних гілок для кожного з наборів є приблизно однаковими. А це є можливим, коли досліджувана складна система є добре структурованою. Припустимо, що справджуються наступні рівності:

$$t_1^M = t_2^M = ... = t_{M_1}^M = t^M;$$

$$t_1^L = t_2^L = ... = t_{M_2}^L = t^L;$$

$$t_1^N = t_2^N = ... = t_R^N = t^N.$$

Тоді формула (14) перетвориться до вигляду (15).

$$S_2 = (M_1 t^M + M_2 t^L + R t^N + t^0)/(t^M + t^L + t^N + t^0). \quad (15)$$

Для найпростішого випадку, коли $M = 3$, $R = 4$, $N_1^0 = N_2^0 = N_3^0 = N_4^0 = 4$, $N = 2$; $L_{r,k} = 2 \; \forall k = \overline{1,2}; r = \overline{1,4}$; $N_r^g = 2 \; \forall r = \overline{1,4}$ одержуємо, що $M_1 = 32$, $M_2 = 16$ та

$$S_2 = (32 t^M + 16 t^L + 4 t^N + t^0)/(t^M + t^L + t^N + t^0) = 11 + (21 t^M + 5 t^L - 7 t^N - 10 t^0)/$$
$$/(t^M + t^L + t^N + t^0). \quad (16)$$

Далі оцінимо $t^M$, $t^L$, $t^N$ та $t^0$, виходячи із наведених вище значень параметрів задачі дослідження складної системи. Унаслідок проведеного аналізу одержуємо, що справджується нерівність $21t^M + 5t^L > 7t^N + 10t^0$, до того ж другий доданок в (16) є меншим за 1, тобто $S_2 > 11$. Отже, навіть для такого простого прикладу системи ми отримуємо суттєве прискорення обчислень, виконуваних за (11). Очевидно, що для більш реального прикладу досліджуваної АТС великого міста величини $M$, $R$, $N_r^0 (r = \overline{1, R})$, $N_r^g (r = \overline{1, R})$, $L_{r,k} (k = \overline{1, N_r^0}; r = \overline{1, R})$ будуть набувати більших значень, тому природньо очікувати, що прискорення $S_2$ буде зростати.

Зазначимо, що агреговане оцінювання стану автошляхів міста та режимів роботи світлофорів також можна здійснювати одночасно, тобто паралельно.

**Розпаралелювання процесу прогностичного оцінювання.** Як було вказано в [7], прогнозування поведінки оцінок $E_M^{r,\,k}$ ($k = \overline{1, N_r^0}$; $r = \overline{1, R}$), $E_M^r$ ($r = \overline{1, R}$) та $E_M$ доцільно використовувати під час планування термінів та обсягів необхідних ремонтних робіт автошляхів міста і пов'язаних з ними витрат. Тому для розпаралелювання процесу прогнозування цих оцінок пропонується використати наступну алгоритмічну конструкцію:

$$fork(h_1^p, h_2^p, \ldots, h_{M_3}^p)\, join, \qquad (17)$$

де $h_{i_4}^p$ ($i_4 = \overline{1, M_3}$) – паралельні автономні гілки, в кожній з яких реалізується процедура прогнозування поведінки однієї з оцінок. Залежно від поставленої задачі оцінювання $M_3$ може приймати різні значення. Наприклад, якщо $M_3 = \sum_{r=1}^{R} N_r^0 + R + 1$, то (17) паралельно реалізує процедуру прогнозування всіх згаданих вище оцінок. Якщо ж $M_3 = \sum_{r=1}^{R} N_r^0 + 1$ або $M_3 = R + 1$, то конструкція (17) паралельно реалізує процедуру прогнозування відповідно оцінок $E_M^{r,\,k}$ ($k = \overline{1, N_r^0}$; $r = \overline{1, R}$), $E_M$ або $E_M^r$ ($r = \overline{1, R}$), $E_M$.

В конструкції (17) складність паралельних гілок є приблизно однаковою, тому можна очікувати, що прискорення обчислень в цьому випадку буде наближатися до свого оптимального значення, тобто до $M_3$. Залежно від доступних обчислювальних ресурсів (кількість процесорів (ядер), кількість обчислювальних вузлів) в кожній гілці можна прогнозувати поведінку певної кількості оцінок.

Зауважимо, що одночасно можна прогнозувати узагальнені оцінки стану АТС (на основі часу проходження елементарних ділянок) та процесу її функціонування (на основі ефективності режиму роботи світлофорів).

**Висновки.** У статті формалізовано методику інтерактивного оцінювання ефективності функціонування АТС великого міста на основі використання методів *U*-статистик. Здійснено оптимізацію (за часом) цієї методики на підставі використання паралельних обчислень. Запропоновано ефективні алгоритмічні конструкції для паралельного виконання локального, агрегованого та прогностичного оцінювання складових АТС. Отримано прискорення для окремих фрагментів паралельних обчислень. Запропоновані алгоритмічні конструкції задають сукупності автономних паралельних гілок, тому вони можуть бути ефективно реалізовані як на обчислювальних засобах зі спільною (комп'ютери з багатоядерним процесором), так і з розподіленою (кластери, гібридні архітектури, високопродуктивні обчислювальні середовища [19–21]) пам'яттю.

Одержані в роботі наукові результати можна використати для оцінювання в режимі реального часу ефективності функціонування АТС великих міст з урахуванням їх характерних особливостей (важливість окремих районів, ділянок шляху, перехресть, маршрутів та їх груп тощо).